# Hierarchical Optimization-Based Model Predictive Control for a Class of Discrete Fuzzy Large-Scale Systems Considering Time-Varying Delays and Disturbances

Mohammad Sarbaz[1], Iman Zamani[1], Mohammad Manthouri[1*], Asier Ibeas[2]

[1]Electrical and Electronic Engineering Department, Shahed University, Tehran, Iran
*Corresponding author. Tel.: +98 21 51212029.
[2]Departament de Telecomunicació i Enginyeria de Sistemes, Escolad'Enginyeria. Universitat Autònoma de Barcelona, Barcelona, Spain
E-mail addresses: mohammad.sarbaz@shahed.ac.ir (M. Sarbaz), zamaniiman@shahed.ac.ir (I. Zamani), mmanthouri@shahed.ac.ir (M. Manthouri), asier.ibeas@uab.cat (A. Ibeas).

*Abstract-* In this manuscript, model predictive control for class of discrete fuzzy large-scale systems subjected to bounded time-varying delay and disturbances is studied. The considered method is Razumikhin for time-varying delay large-scale systems, in which it includes a Lyapunov function associated with the original non-augmented state space of system dynamics in comparison with the Krasovskii method. As a rule, the Razumikhin method has a perfect potential to avoid the inherent complexity of the Krasovskii method especially in the presence of large delays and disturbances. The considered large-scale system in this manuscript is decomposed into several subsystems, each of which is represented by a fuzzy Takagi-Sugeno (T-S) model and the interconnection between any two subsystems is considered. Because the main section of the model predictive control is optimization, the hierarchical scheme is performed for the optimization problem. Furthermore, persistent disturbances are considered that robust positive invariance and input-to-state stability under such circumstances are studied. The linear matrix inequalities (LMIs) method is performed for our computations. So the closed-loop large-scale system is asymptotically stable. Ultimately, by two examples, the effectiveness of the proposed method is illustrated, and a comparison with other papers is made by remarks.

## I. INTRODUCTION

Large-scale systems have become interesting in control engineering. Many systems become large in scope and complex in scale since decades ago and coping with these systems requires time and efforts since many problems are appeared like existence of disturbances, nonlinear parameters and complex dynamic. Many algorithms have proposed for large-scale systems since years ago [1] and [2]. The decentralized controller is an interesting approach that scholars aimed to solve these problems [3]. In this approach, the considered large-scale system is decomposed into several subsystems and a controller is applied to each subsystem separately, and the interconnections between subsystems are considered. So the overall system is stabilized by adopting this approach. In [4], a decentralized fault-tolerant tracking control is applied to a large-scale system with sensor and actuator faults that its problem is decomposed into several problems. The decentralized adaptive fuzzy controller is proposed for a class of uncertain nonlinear large-scale systems considering random sensor delays and random sensor nonlinearities in [5]. By using a group of sensor blocks two new distributed data-driven optimal fault detection approaches are applied to a large-scale system in [6].

As it was mentioned above, one of the important issues in large-scale system is its complex dynamic. To design a suitable controller and solve problems, it is significant to have the exact model of the system. But it is almost impossible when the system has many nonlinear parameters and severe dynamic. In this case, scientists use some methods like fuzzy logic or neural networks to identify the model of the system [7] and [8]. For instance, in [9], the neural network is used to estimate the inverse dynamic of the da Vinci surgical robot to enables estimation of the external environment forces. In this paper, since the considered system is large-scale, the fuzzy model of the systems is used instead of the dynamic model. Nowadays, fuzzy systems with IF-THEN rules have become popular and useful approach for systems with large, complex, and nonlinear relations and several researches have been done [10], [11], and [12]. [13] Introduces a type of fuzzy inference system well-known Takagi-Sugeno, fuzzy model, and model complex nonlinear system to arbitrary degrees of accuracy. On the other hand, another vital issue that is considered in this paper is time-varying delay that is unavoidable problem. The time delay phenomena can be seen in many real and industrial systems [14], and in many cases it causes instability and might lead outputs to unexpected goal. Since the time delay causes instability, attentions have been attracted for many years. Totally, Lyapunov theory for systems with time-varying delay can be proposed into two approaches, the Lyapunov-Krasovskii functional (LKF) and Lyapunov-Razumikhin function (LRF) [15]. For discrete time systems, the Krasovskii method makes use of an augmentation of the state vector with all delayed states, which yields the application of classical Lyapunov methods to an augmented system without delay. In the Krasovskii method, an augmentation of the state vector with all delayed states are used, which yields the applications of classical Lyapunov methods to an augmented system without delay. So the computational burden in this method will be heavy, especially for large-scale systems [16]. On the contrast, the Razumikhin method includes a Lyapunov function for the original non-augmented system despite of the fact that it might be a bit conservative. Thereby, Razumikhin method has a better potential to avoid the complexity of the Krasovskii method. Many studies have been done by adopting Razumikhin method [17], [18], and [19]. In [16], the author proposed an LMI-based model predictive control for fuzzy T-S system with time-varying delay and used Razumikhin method for the delay problem and the LMI-based method is proposed for not interconnected system. In [20], a model predictive control strategy for nonlinear time-delay systems with unknown time varying delay is considered instead of the traditional constant delay. Nonlinear discrete time-delay systems are represented by T–S fuzzy systems comprised of piecewise linear delay difference equations in [21]. The stability of delay coupled systems with hybrid switching diffusions based on Razumikhin method is studied in [22] and it combines Lyapunov method with graph theory for typical systems. To stabilize these systems that are involved time-varying delay, many control systems have been applied like adaptive, fuzzy, robust, or so. One of the most interesting controller has been model predictive control.

Model predictive control (MPC) has attended popularity as a reliable control approach. By properly using the system model to predict the output response, MPC methods allow to choose the optimal control action that minimizes a desired cost function. In many applications, MPC has been recognized as a viable alternative to many other classic schemes based. For such applications, the advantages introduced by MPC are the higher dynamic performance, the possibility of performing a multivariable controller design, the inclusions of constraints on input and output variables, and the possibility of including nonlinearities in both the model and the constraints. Although the need for an accurate model may represent a drawback in some applications, it is remarked how reliable model of systems are usually available for control design [23] and [24]. The gist of the model predictive control is an optimal control sequence, which is computed by minimizing a finite horizon cost function at each sampling time. Studying on nonlinear robust model predictive control for Takagi-Sugeno fuzzy systems can be seen in [25]. As mentioned in [26] and [27], invariant set theory can provide sufficient conditions to make sure the recursive feasibility and the closed-loop system stability. In [28] both online and off-line robust fuzzy model predictive control with structured uncertainties and persistent disturbances is studied. In [23] robust fuzzy model predictive control with nonlinear local models is introduced, in which, for both nonlinear and linear states of system, separated controllers are designed. In [29], an LMI-based MPC is proposed for a large-scale system with disturbances and uncertainties. Obviously, it can be seen that hierarchical-based optimization MPC for large-scale systems with time-varying delay, persistent disturbances, and with respect to Lyapunov-Razumikhin function has not been studied yet and several problems remain unsolved. In this study, the considered controller is model predictive control. Since the considered system is a large-scale, the hierarchical scheme is applied to solve the optimization problem based on decomposition and coordination concept [30]. On the other hand, no research has been done regarding the MPC for fuzzy large-scale systems with time-varying delay adopting the Lyapunov-Razumikhin function. Therefore, the MPC is applied to a large-scale system that its dynamic is modeled by T-S fuzzy, and the Lyapunov-Razumikhin function is considered for the time-varying delay problem. Besides, the hierarchical scheme is assumed for the optimization problem of the system.

So the contribution of this paper can be summarized as:
- ➢ Proposing the fuzzy model predictive control for discrete nonlinear large-scale system.

- Considering the time-varying delay and persistent disturbances simultaneously.
- Applying hierarchical optimization to the considered large-scale system.
- Studying Lyapunov-Razumikhin function.
- Analyzing the input-to-state stability.
- Using the fuzzy Takagi-Sugeno model of the large-scale system.
- Inspiring the $H_\infty$ performance to encounter the effectiveness of disturbances.

The remainder of this paper is: in Section II, some preliminaries information about fuzzy model predictive control of Takagi-Sugeno large-scale systems with time-varying delay and hierarchical optimization scheme are introduced. In Section III, robust positive invariance (RPI), the computation of terminal constraint set for nonlinear model-based fuzzy systems is provided. In Section IV, two numerical examples are illustrated. Finally, the conclusion is given in Section V.

Notations: $R, R_+, Z$, and $Z_+$ depict sets of real numbers, positive real numbers, integers, and positive integers, respectively. $Z_{[m,n]}$ is Symbol of set of integers in the interval $[m,n]$ for convenience. $\|x\|$ depicts the norm of vector $x_i \in R^c$. A real-valued scalar function $\partial: R_+ \to R_+$ is an $\mathcal{H}$-function ($\partial \in \mathcal{H}$), if it is continuous, rigidly increasing and $\partial(0) = 0$. So we can say, $\partial \in \mathcal{H}_\infty$ if $\partial \in \mathcal{H}$ and $\lim_{s \to \infty} \partial(s) = \infty$. A function $\beta: R_+ \times R_+ \to R_+$ is an $\mathcal{H}L$-function ($\beta \in \mathcal{H}L$), if for each fixed $s > 0, \beta(.,s)$ is a $\mathcal{H}$-function, and for each fixed $r > 0, \beta(r,.)$ is rigidly decreasing and $\beta(r,s) \to 0$ as $s \to \infty$.

## II. PRELIMINARIES

### A. System Description

The following discrete-time nonlinear large-scale system is assumed:

$$x(k+1) \in f\big(x_i(k), x_{id}(k), u_i(k), d_i(k)\big) \quad k \in Z_+ \tag{1}$$

where $x_i(k), x_{id}(k) \in \mathbb{X}, u_i(k) \in \mathbb{U}$ and $d_i(k) \in \mathbb{D}$, with $\mathbb{X} \subseteq \mathbb{R}^n, \mathbb{U} \subseteq \mathbb{R}^m$, and $\mathbb{D} \subseteq \mathbb{R}^c$, are system current and delayed states, inputs, and disturbances, respectively. $x_{id}(k)$ is defined as follows:

$$x_{id}(k) = x_{id}(k + d(k)), \quad d(k) \in \mathbb{Z}_{[-h,-1]}. \tag{2}$$

where $d(k)$ denotes the number of delay, h denotes the upper bound of delay, and the minimal delay is set as 1. Now, for time-delay systems some definitions related to the robust positive invariance and input-to-state stability are proposed [31]:

*Definition 1(Robust Positively Invariant (RPI) set):* for the Lyapunov-Razumikhin function conditions, consider system (1), a set $\Omega_{is}$ is an RPI set for the closed-loop system corresponding to the control law, and for the $d(k) \in \mathbb{Z}_{[-h,-1]}$, if $\forall x_i, x_{id} \in \Omega_{is}$, and $\forall d_i(k) \in \mathbb{D}$, the control effort assures that $x_i^+ \in \Omega_{is}$.

*Definition 2 (Input-to-state-Stability (ISS)):* A discrete-time nonlinear system $x(k+1) = f(x_i(k), x_{id}(k), d_i(k))$ where $d$ symbolizes the disturbance vector, is input-to-state stable (ISS) if there exists an $\mathcal{H}L$-function $\beta$ and a $\mathcal{H}$-function $\gamma$ such that for each input $d$, $\|x(k)\| \leq \beta\left(\left\|x_{i_{[-h,0]}}\right\|, k\right) + \gamma\left(\| d_{i_{[0,k-1]}} \|\right)$, where $x_{i_{[-h,0]}} \in \mathbb{X}^{h+1}$ is the initial (delayed) state vector, $d_{i_{[0,k-1]}} \in \mathbb{W}^k$ is the disturbance sequence, $k \in Z_+$.

*Definition 3 (ISS Lyapunov-Razumikhin Function [16])* A continuous positive definite function $V(x(k))$ is called an ISS-Lyapunov-Razumikhin function for system $x(k+1) = f(x_i(k), x_{id}(k), d_i(k))$, if there exist $\mathcal{H}_\infty$-function $\partial_1, \partial_2$, and $\mathcal{H}$-function $\rho$ such that:

$$\partial_1(\| x(k) \|) \leq V(x(k)) \leq \partial_2(\| x(k) \|) \tag{3}$$

$$V(x(k+1)) \leq max\{\bar{V}(x(k)), \rho(\| d(k) \|)\} \tag{4}$$

where $\bar{V}(x(k)) = max\{V(x_i(k)), V(x_{id}(k))\}$

*Lemma 1[32]:* If system $x(k+1) = f(x_i(k), x_{id}(k), d_i(k))$ admits an ISS-Lyapunov- Razumikhin function, then it is ISS.

### B. Time delay Takagi-Sugeno system description

The fuzzy Takagi-Sugeno large-scale system composed of $N$ subsystems with time-varying delay is:

$$S_i^l: \begin{cases} \text{IF } z_{i1} \text{ is } F_{i1}^l \text{ and } \dots \text{ and } z_{ig} \text{ is } F_{ig}^l \\ \text{THEN } x_i(k+1) = A_i^l x_i(k) + B_i^l u_i(k) + A_{id}^l x_{id}(k) + w_i^l d_i(k) + \sum_{\substack{j=1 \\ i \neq j}}^{N} f_{ij} x_j(k) \end{cases} \quad (5)$$

in which $A_i^l, B_i^l$, and $w_i^l$ $(i = 1,2,\dots,N; l = 1,2,\dots,r_i)$ are the system matrices and disturbances of rule-$l$ in subsystem $S_i$. Here, $x_i(k) \in R^c$, $u_i(k) \in R^n$, and $d_i(k) \in R^m$ are state vector, input vector, and disturbance vector. $r_i$, $f_{ij}, z_i(k) = [z_{i1}, z_{i2}, \dots, z_{ig}]$, and $F_{iq}^l (q = 1,2,\dots,g)$, respectively, illustrate the number of the fuzzy rules in subsystem $S_i$, the interconnection between subsystem $S_i$ and $S_j$, some measurable premise variables, and the linguistic fuzzy sets of the rule $l$. The fuzzy large-scale system (5) with time-varying delay can be shown as:

$$x_i^+ = A_{i\mu} x_i + B_{i\mu} u_i + A_{id\mu} x_{id} + w_{i\mu} d_i + \sum_{\substack{j=1 \\ i \neq j}}^{N} f_{ij} x_j, \quad i = 1,2,\dots,N \quad (6)$$

$$A_{i\mu} = \sum_{l=1}^{r_i} \mu_i^l(z_{iq}) A_i^l \quad ; \quad A_{id\mu} = \sum_{l=1}^{r_i} \mu_i^l(z_{iq}) A_{id}^l \quad ; \quad B_{i\mu} = \sum_{l=1}^{r_i} \mu_i^l(z_{iq}) B_i^l$$

$$w_{i\mu} = \sum_{l=1}^{r_i} \mu_i^l(z_{iq}) w_i^l \quad ; \quad f_{ij\mu} = \sum_{l=1}^{r_i} \mu_i^l(z_{iq}) f_{ij} \quad (7)$$

and $\mu_i^l(z_{iq})$ represents the normalized membership function.

Control law is shown as:

$$C_i^l: \begin{cases} \text{IF } z_{il} \text{ is } F_{i1}^l \text{ and } \dots \text{ and } z_{ig} \text{ is } F_{ig}^l \\ \text{THEN } u_i(k) = k_i^l x_i(k) \end{cases} \quad (8)$$

With the same weight notation $\mu_i^l(z_{iq})$ and respect to (8), the final output of the controller will be:

$$u_i(k) = \sum_{l=1}^{r_i} \mu_i^l(z_{iq}) k_i^l x_i(k) \quad (9)$$

The closed-loop system with time-varying delay will be:

$$x_i(k+1) = \sum_{l=1}^{r_i} \sum_{m=1}^{r_i} \mu_i^l(z_{iq}) \mu_i^m(z_{iq}) \left( [A_i^l + B_i^l k_i^m] x_i(k) + A_{id}^l x_{id}(k) + w_i^l d_i(k) \right) + \sum_{l=1}^{r_i} \sum_{\substack{j=1 \\ i \neq j}}^{N} \mu_i^l(z_{iq}) f_{ij} x_j(k) \quad (10)$$

## C. Model Predictive Control

The prediction model and cost function are proposed here. The prediction form is:

$$x_i(k+n+1|k) = A_{i\mu} x_i(k+t|k) + A_{id\mu} x_{id}(k+t|k) + B_{i\mu} u_i(k+t|k) + w_{i\mu} d_i(k+t|k)$$
$$+ \sum_{\substack{j=1 \\ i \neq j}}^{N} f_{ij\mu} x_j(k+t|k) \quad (11)$$

and the cost function is:

$$J(k) = \sum_{i=1}^{N} j_i(k) = \sum_{i=1}^{N} \left( \Pi_i(K) + \sum_{n=0}^{T-1} \Pi_i(k+t|k) + V_{it}(x_i(k+T|k)) \right) \quad (12)$$

where $\Pi_i(k+n|k)$ and $V_{it}(x_i(k+T|k))$ are stage cost at the predicted time instant and terminal cost, and $T$ is the prediction length. It is highly notable that $V_{it}(\cdot)$ must be a positive function [33], and the stage cost is chosen as:

$$\Pi(k) = \sum_{i=1}^{N} \Pi_i(k)$$

$$= \sum_{i=1}^{N} \left( x_i^T(k+t|k) Q x_i(k+t|k) + u_i^T(k+t|k) R u_i(k+t|k) - \tau_i d_i^T(k+t|k) d_i(k+t|k) \right) \quad (13)$$

where $R$ and $Q$ are real fixed matrices, and $\tau_i$ is a positive scalar. It is evident the cost function includes the disturbance, and is influenced by the $H_\infty$ control [34]. Consequently, the cost function cannot be optimized directly as the disturbance is included. Instead, a min-max technique is chosen that minimizing the worst-case cost function [35]. Furthermore, at the end of the prediction, it is mostly required that the states enter a terminal constraint set to attain asymptotic stability, since it is almost impossible task to achieve the asymptotic stability to the origin, in the presence of persistent disturbance [36]. $x_i(k+T|k) \in \Omega_{is}$ shows the terminal constraint set. The online optimization problem is:

$$\min_{u_i(k+t|k)} \max_{d_i(k+t|k)} J_i(k),$$
$$s.t. \quad u_i(k+t|k) \in U_i$$
$$d_i(k+t|k) \in D_i$$
$$x_i(k+T|k) \in \Omega_{is}$$

where $t \in Z_{[1,T-1]}$ and $\Omega_{is}$ shows the terminal constraint set. Here $d_i \in D_i \coloneqq \{d_i | d_i^T d_i \leq \gamma_i^2\}$, $u_i \in U_i \coloneqq \{u_i | |u_{im}| \leq u_{im.max}\}$ and $\gamma_i^2$ is a positive scalar, $u_{im}$ is the $m$-th element of the inputs, $m \in Z_{[1,w]}$.

As it was mentioned, the model predictive control is a kind of controller which computes the input vector by minimizing a specified cost function. In this manuscript, the considered system is large-scale and the form of the controller is decentralized. Thus, to avoid the uncertainty in cost function and reach the best solution, the hierarchical scheme is applied to the optimization problem [37].

### III. MAIN RESULTS

#### A. RPI Set for Fuzzy Takagi-Sugeno Large-Scale System with Time-Varying Delay

First, the RPI property and terminal constraint set are defined. Second, the recursive feasibility is analyzed. Finally, the ISS will be provided. The RPI set is illustrated by $\Omega_{is}$, and $\Lambda_i(x)$ as the corresponding control law. The RPI set is defined as, $\forall x_i, x_{id} \in \Omega_{is}$, the control effort assures that $x_i^+ \in \Omega_{is}$, for all allowable disturbance. For the fuzzy Takagi-Sugeno large-scale system (3) with time-varying delay, define $\Omega_{is}$ as:

$$\Omega_{is} \coloneqq \left\{ \{x_i, x_{id}\} \middle| \max \left\{ \left( \sum_{l=1}^{r_i} \mu_i^l(z_{iq}) x_i^T P_{i\mu} x_i \right), \left( \sum_{l=1}^{r_i} \mu_i^l(z_{iq}) x_{id}^T P_{i\mu} x_{id} \right) \right\} \leq \varsigma_i \right\} \quad (14)$$

where $P_{i\mu} = \sum_{l=1}^{r_i} \mu_i^l(z_{iq}) P_i$ and $\varsigma_i$ is a positive scalar and the corresponding control law is $\Lambda_i(x_i, x_{id}) = \sum_{l=1}^{r_i} \mu_i^l(z_{iq}) k_i^l x_i(k)$.

*Lemma 2 [28]:* The set $\Omega_{is}$ is an RPI set if there exists a positive scalar $\lambda_i$, $(0 < \lambda_i < 1)$, such that,

$$\sum_{i=1}^{N} \left\{ \frac{1}{\varsigma_i} x_i^{+T} P_{i\mu}^+ x_i^+ - \frac{1-\lambda_i}{\varsigma_i} \max\{x_i^T P_{i\mu} x_i, x_{id}^T P_{i\mu} x_{id}\} - \frac{\lambda_i}{\gamma_i^2} d_i^T d_i \right\} \leq 0 \quad (15)$$

with $P_{i\mu}^+ = \sum_{l=1}^{r_i} \mu_i^l(x_i^+) P_i$, for all $x_i^+ \in A_{i\mu} x_i + A_{id\mu} x_{id} + B_{i\mu} u_i + w_{i\mu} d_i + \sum_{\substack{j=1 \\ i \neq j}}^{N} f_{ij\mu} x_j$, $u_i \in U_i$, and $d_i \in D_i$.

*Remark 1:* Here, by proposing the *Theorem 1*, two LMIs are solved and if they are feasible, two results are achieved. (1) It will be proved that the considered large-scale fuzzy system with time-varying delay is stable in the sense of Lyapunov. (2) Controller's gains are computed in the restricted bound, optimally.

*Theorem 1:* Consider the fuzzy system (5), if there exist positive definite matrices $X_i$ $X_j$, and $X_{i\mu}$, positive scalar $\lambda_i$ $(0 < \lambda_i < 1)$ such that the following matrix inequalities are met:

$$\begin{bmatrix} w_{i\mu}^T X_i w_{i\mu} - \varsigma_i \lambda_i \varpi_i & \star & & \star \\ \Theta_i^T X_i w_{i\mu} & N\sqrt{a} \sum_{\substack{j=1 \\ i \neq j}}^{N} f_{ij}^T X_j f_{ij} - \varrho_i(-\lambda_i + 1)X_{i\mu} & & \star \\ A_{id\mu}^T X_i w_{i\mu} & A_{id\mu}^T X_i \Theta_i & A_{id\mu}^T X_i A_{id\mu} - (-\lambda_i + 1)\varrho_{id} X_{i\mu} \\ f_{ij}^T X_i w_{i\mu} & (1-\sqrt{\alpha})f_{ij}^T X_i \Theta_i & f_{ij}^T X_i A_{id\mu} \\ \vdots & \vdots & \vdots \\ f_{iN}^T X_i w_{i\mu} & (1-\sqrt{\alpha})f_{iN}^T X_i \Theta_i & f_{iN}^T X_i A_{id\mu} \\ 0 & X_i \Theta_i & 0 \\ \star & \cdots & \star & \star \\ \star & \cdots & \star & \star \\ \star & \cdots & \star & \star \\ -(\alpha-1)f_{ij}^T X_i f_{ij} & \cdots & \star & \star \\ \vdots & \ddots & \vdots & \vdots \\ -(\alpha-1)f_{iN}^T X_i f_{ij} & \cdots & -(\alpha-1)f_{iN}^T X_i f_{iN} & \star \\ 0 & \cdots & 0 & -N^{-1}X_i^T \end{bmatrix} < 0 \quad (16)$$

$$\begin{bmatrix} Z_i & k_{i\mu}^T \\ k_{i\mu} & 1 \end{bmatrix} \geq 0 \quad Z_{iss} \leq u_{is,max}^2, s \in Z_{[1,m]} \quad (17)$$

then the set $\Omega_{is} := \left\{ \{x_i, x_{id}\} \middle| max\left\{ \left(\sum_{l=1}^{r_i} \mu_i^l(z_{iq}) x_i^T P_{i\mu} x_i\right), \left(\sum_{l=1}^{r_i} \mu_i^l(z_{iq}) x_{id}^T P_{i\mu} x_{id}\right) \right\} \leq \varsigma_i \right\}$ is a RPI set for the fuzzy system (5) corresponding to the feedback control law $\Lambda_i(x_i, x_{id}) = \sum_{l=1}^{r_i} \mu_i^l(z_{iq}) k_i^l x_i(k)$. Where $\theta_i = (A_{i\mu} + B_{i\mu} k_{i\mu}), Z_{iww}$ is $w-th$ diagonal element of matrix $Z_i$, $N$ represents the number of subsystems, $\varrho_i$ and $\varrho_{id}$ are fixed positive values which satisfy $\varrho_i + \varrho_{id} = 1, X_i = \varsigma_i P_i, X_{i\mu} = \varsigma_i P_{i\mu}, X_j = \varsigma_i P_j, \varpi_i = \frac{\varsigma_i}{\gamma_i^2}, \alpha \geq 2$, and $i, j, l, N, \alpha \in R_+$.

*Proof:* See Appendix **A**

## B. The coordination sub-network

In this section the hierarchical optimization will be carried on. For this, the Hamiltonian function is defined as:

$$H(\cdot) = \sum_{i=1}^{N} H_i(\cdot) = \sum_{i=1}^{N} \Bigg\{ \Pi_i(K) \\ + \sum_{k=0}^{K-1} \Bigg( x_i^T(k)Qx_i(k) + u_i^T(k)Ru_i(k) - \tau_i d_i^T(k)d_i(k) + \delta_i^T(k)\left(z_i(k) - \sum_{\substack{j=1 \\ i \neq j}}^{N} f_{ij}x_j(k)\right) \\ + \bar{p}_i^T(k+1)\left(-x_i(k+1) + g_i(x_i(k), u_i(k), z_i(k))\right) \Bigg) \Bigg\} \quad (18)$$

where $g_i(x_i(k), u_i(k), z_i(k)) = A_{i\mu} x_i + A_{id\mu} x_{id} + B_{i\mu} u_i + w_{i\mu} d_i + C_i z_i(k)$ and $\Pi_i(K)$ is continuously differentiable.

here, $\delta_i$, $\bar{p}_i$, and $C_i$ are the Hamiltonian multipliers, the co-state variables, and fixed value matrix. By the interaction prediction strategy, the coordination for hierarchical optimization is to find an approach for updating the coordination values $\delta_i$ and $z_i(k)$:

$$\frac{\partial H(\cdot)}{\partial z_i(k)} = 0, \quad \frac{\partial H(\cdot)}{\partial \delta_i(k)} = 0 \quad i = 1, \dots, N \,;\; k = 0, \dots K-1 \tag{19}$$

thereby,

$$\delta_i(k) = -\frac{\partial g_i(x_i, u_i, z_i)}{\partial z_i}\bar{p}_i(k+1) = -C_i^T \bar{p}_i(k+1), \quad i = 1, \dots, N \,;\; k = 0, \dots K-1 \tag{20}$$

$$z_i(k) = \sum_{\substack{j=1 \\ i \neq j}}^{N} f_{ij} x_j(k) \quad i = 1, \dots, N \,;\; k = 0, \dots K-1 \tag{21}$$

## C. The local optimization sub-networks

In local sub-networks, exchanging the information between the coordination and local sub-networks will make up the hierarchical scheme as shown in Fig. 1. In this figure $\delta_i = \sigma_i$.

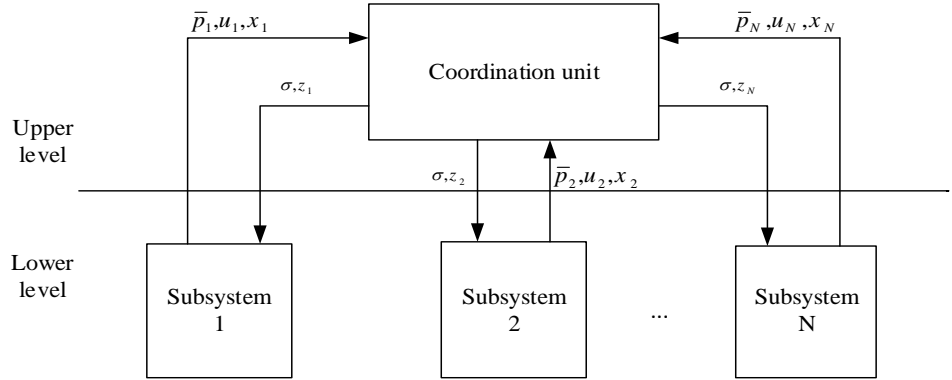

Fig. 1 Architectures of interactions

in the final step of hierarchical optimization, the convergence must be checked by evaluating the overall interaction error:

$$e(k) = \sum_{i=1}^{N} \sum_{k=1}^{K-1} \left\{ z_i(k) - \sum_{\substack{j=1 \\ i \neq j}}^{N} f_{ij} x_j(k) \right\}^T \left\{ z_i(k) - \sum_{\substack{j=1 \\ i \neq j}}^{N} f_{ij} x_j(k) \right\} \tag{22}$$

now, provided the desire convergence obtained, stop. Otherwise, repeat the steps again.

## D. The terminal constraint set

The terminal constraint set $\Omega_{is}$ should satisfy two conditions. First, it should be RPI set. Second, a positive definite function (terminal cost function) $V_i(x)$ should exist such that $\forall\, x_i \in \Omega_{is}$,

$$\gamma_3(\| x_i \|) \leq V(x_i) \leq \gamma_4(\| x_i \|) \tag{23}$$

$$\sum_{i=1}^{N} V(x_i^+) - V(x_i) < -\sum_{i=1}^{N} H_i(.) \tag{24}$$

where $\varsigma_3$ and $\varsigma_4$ are $H_\infty$ functions, $V(x)$ is given as:

$$V(x) = \sum_{i=1}^{N} V_i(x) = \sum_{i=1}^{N} \sum_{l=1}^{r_i} \mu_i^l(z_{iq}) x_i^T P_i x_i \tag{25}$$

*Remark 2*: In this section, by proposing the *Theorem 2*, the concept of the robust positively invariant and constraint set are achieved through solving a LMI, and it is ensured that trajectories of the large-scale system with time-varying delay are stable robustly.

For convenience, a table of notations is provided:

Table 1. Abbreviations and notation

| notation | definition | notation | definition |
|---|---|---|---|
| $A_i^l, B_i^l,$ and $w_i^l$ | System matrices and disturbances | $R$ and $Q$ | Positive weights |
| $X_i$ | $X_i = \xi_i P_i$ | $H_i$ | $\varsigma_i R$ |
| $X_j$ | $X_j = \xi_i P_j$ | $N$ | Number of subsystems |
| $P_i$ | Positive matrix | $\alpha$ | Positive scalar |
| $f_{ij}$ | Interconnection between subsystem | $\gamma_i^2$ | Positive scalar |
| $\varsigma_i$ | Positive scalar variable | $\varpi_i$ | $\frac{\varsigma_i}{\gamma_i^2}$ |
| $\tau_i$ | Positive scalar | $\varrho_i$ | Positive value |
| $\lambda_i$ | Positive scalar | $\delta_i$ | Hamiltonian multiplier |
| $\bar{X}_i^T(k+1)$ | $\varsigma_i \bar{p}_i^T(k+1)$ | $\bar{p}_i(k+1)$ | Co-state variable |
| $e(k)$ | Overall interaction error | $C_i$ | Fixed value matrix |

*Theorem 2*: Consider the fuzzy Takagi-Sugeno system (5), if (16), (17) and the following matrix inequality is feasible,

$$\begin{bmatrix}
w_{i\mu}^T X_i w_{i\mu} - \varsigma_i \tau_i & \star & \star \\
\theta_i^T X_i w_{i\mu} & N\sqrt{a} \sum_{\substack{j=1 \\ i \neq j}}^{N} f_{ij}^T X_j f_{ij} - \varrho_i X_{i\mu} + \varsigma_i Q & \star \\
A_{id\mu}^T X_i w_{i\mu} & A_{id\mu}^T X_i \theta_i & A_{id\mu}^T X_i A_{id\mu} - \varrho_{id} X_{i\mu} \\
f_{ij}^T X_i w_{i\mu} & (1-\sqrt{\alpha}) f_{ij}^T X_i \theta_i & f_{ij}^T X_i A_{id\mu} \\
\vdots & \vdots & \vdots \\
f_{iN}^T X_i w_{i\mu} & (1-\sqrt{\alpha}) f_{iN}^T X_i \theta_i & f_{iN}^T X_i A_{id\mu} \\
0 & 0 & 0 \\
0 & H_i k_{i\mu} & 0 \\
0 & X_i \theta_i & 0
\end{bmatrix}$$

$$\begin{matrix}
\star & \cdots & \star & \star & \star & \star \\
\star & \cdots & \star & \star & \star & \star \\
\star & \cdots & \star & \star & \star & \star \\
-(\alpha-1) f_{ij}^T X_i f_{ij} & \cdots & \star & \star & \star & \star \\
\vdots & \ddots & \vdots & \vdots & \vdots & \vdots \\
-(\alpha-1) f_{iN}^T X_i f_{ij} & \cdots & -(\alpha-1) f_{iN}^T X_i f_{iN} & \star & \star & \star \\
-\frac{1}{2}\left(\bar{X}_i^T(k+1) + \varsigma_i \delta_j^T(k)\right) f_{ij} & \cdots & -\frac{1}{2}\left(\bar{X}_i^T(k+1) + \varsigma_i \delta_N^T(k)\right) f_{iN} & \left(\varsigma_i \delta_i^T(k) z_i(k) + \bar{X}_i^T(k+1) z_i(k)\right) & \star & \star \\
0 & \cdots & 0 & 0 & -H_i & 0 \\
0 & \cdots & 0 & 0 & 0 & -N^{-1} X_i^T
\end{matrix}$$

$$< 0 \tag{26}$$

then $\Omega_{is}$ is a terminal constraint set considering the terminal cost function. Where $N$ represents the number of subsystems, $\theta_i = (A_{i\mu} + B_{i\mu}k_{i\mu})$, $\bar{X}_i^T(k) = \varsigma_i \bar{p}_i^T(k)$, $H_i = \varsigma_i R$, $X_i = \varsigma_i P_i$, $X_{i\mu} = \varsigma_i P_{i\mu}$, $X_j = \varsigma_i P_j$, $\alpha \geq 2$, and $i, j, l, N, \alpha \in R_+$.

*Proof:* See Appendix **B**

### E. Control Algorithm

Based on the recent results and to finalize the control design, now, the online control algorithm is studied. Therefore, the terminal constraint set $V(x(k))$, see (24), should satisfy the following condition,

$$V(x(k)) = \sum_{i=1}^{N} \bar{V}_i(x(k)) = \sum_{i=1}^{N} \sum_{l=1}^{r_i} \mu_i^l(z_{iq}) x_i^T(k+d) P_i x_i(k+d) \leq \varsigma_i \quad ; \quad d(k) \in \mathbb{Z}_{[-h,-1]}. \tag{27}$$

the following optimization problem is for minimizing $\varsigma_i$:

$$\min \varsigma_i \text{ subject to } \bar{V}_i(x(k)) \leq \varsigma_i \tag{28}$$

furthermore, a proper condition for $\bar{V}_i(x(k)) \leq \varsigma_i$ is:

$$x_i^T(k+d) P_i x_i(k+d) \leq \xi_i \quad ; \quad i = 1, \ldots, N$$

which is $\varsigma_i - x_i^T(k+d) P_i x_i(k+d) \geq 0$ and equal to $\varsigma_i - x_i^T(k+d) \frac{\varsigma_i P_i}{\varsigma_i} x_i(k+d) \geq 0$

by defining symmetrical matrix $X_i = \varsigma_i P_i$, is guaranteed by the following LMIs,

$$\begin{bmatrix} \varsigma_i & x_i^T(k+d) \\ x_i(k+d) & X_i^{-1}\varsigma_i \end{bmatrix} \geq 0 \quad ; \quad d(k) \in \mathbb{Z}_{[-h,-1]}. \tag{29}$$

Note 1: The symmetric matrix $X_i$ is defined as a matrix with an achievable inversion. So the inversion of $X_i$ must be considered in (29).

Note 2: Since the first Example is a mathematical Example, the interconnection between two subsystems are considered as mathematical matrices. But in a real example, the interconnection between two subsystems must be modeled while the engineer is modeling the dynamic of the system. Interconnections in a large-scale system are same as system matrices like $A_i^l$ and $B_i^l$. These are computed when the system's dynamic is linearized.

*Remark 3:* As mentioned in introductory section, an important issue in MPC is optimization problem. The considered system in this paper is large-scale and values of the LMIs are computed and controlled the closed-loop system based on the minimized value of $\varsigma_i$. In the previous *theorems*, the bilinear matrix inequalities have been proposed. Since the significant parts of this paper are gains and minimized $\varsigma_i$, the values of $X_i, X_{i\mu}, \lambda_i, \varpi_i$, and $H_i$ are defined until the feasible solutions for gains and minimized value of $\varsigma_i$.

**Algorithm**

**Step 1 (Coordination optimization):** Set initial values for $\delta_i(k), z_i(k)$, and different values of $\varpi_i, H_i, X_j, X_i$, and $\lambda_i$.

**Step 2**: Solve the following optimization problem

$$\min_{h_{i\mu}, X_i, z_i} \varsigma_i, \tag{30}$$

subject to (16), (17), (26), (29),

and for each subsystem, find the values of $\varsigma_i$, $k_{i\mu}, \bar{X}_i^T(k+1) = \varsigma_i \bar{p}_i^T(k+1), \delta_i(k) = -C_i \bar{p}_i(k+1), X_{i\mu} = \varsigma_i P_{i\mu}, X_i = \varsigma_i P_i, \varpi_i = \frac{\varsigma_i}{\gamma_i^2}, H_i = R\varsigma_i$.

**Step 3:** Check for the converging by the overall interaction error.

$$e(k) = \sum_{i=1}^{N}\sum_{k=1}^{K-1}\left\{z_i(k) - \sum_{\substack{j=1\\i\neq j}}^{N} f_{ij}x_j(k)\right\}^T \left\{z_i(k) - \sum_{\substack{j=1\\i\neq j}}^{N} f_{ij}x_j(k)\right\}$$

**Step 4:** If desired values of coordination achieved, stop. Otherwise, Set $\begin{bmatrix}\delta_i(k)\\z_i(k)\end{bmatrix}^{l+1} = \begin{bmatrix}-C_i\bar{p}_i(k+1)\\\sum_{\substack{j=1\\i\neq j}}^{N} f_{ij}x_j(k)\end{bmatrix}^l$ and go **step 2.**

One of the significant part of the model predictive control is recursive feasibility. In the following *theorem*, it will be proven that the recursive feasibility can be achieved and the constrained optimization problem has the solution at any time.

*Theorem 3:* for the system (5), it will always be solvable if the solution of the optimization problem can be achievable at time 0, and the recursive feasibility can be obtained.

*Proof:* If (30) is feasible at time $k$, then resorting to the $x_i, x_{id} \in \Omega_{is}$, and according to the theorem 1, which has been implied that $x_i(k+1) \in \Omega_{is}$, then it can be concluded that (30) can be solvable at time $k+1$. Besides, the solution which achieved at the time $k$, is feasible at time $k+1$, and also the optimization problem is feasible at all time.

*Theorem 4*: should the constrained optimization problem is feasible at the initial time 0, system (5) is **ISS** due to the disturbance $d$.

*Proof:* Assume the Lyapunov-Razumikhin function $V(x_i(k)) = x_i^T(k)P^*_{i\mu}x_i(k)$, consider that it gets its maximum values at the delayed states $x_i(k+\Delta)$, $\Delta \in \{d(k), 0\}$, and $d(k)$ defined in (2). $\bar{V}(x_i(k)) = x_i^T(k+\Delta)P^*_{i\mu}x_i(k+\Delta)$, in which $P^*_{i\mu} = \sum_{l=1}^{r_i}\mu_i^l(z_{iq})P_i^*$, here, $P_i^*(k)$ is an optimal value of $P_i(k)$ at time $k$, then it can be obtained:

$$\psi^*_{min}\|x_i(k)\|^2 \leq V(k, x_i) \leq \psi^*_{max}\|x_i(k)\|^2 \tag{31}$$

in which
$$\psi^*_{max} = max\{\psi_{max}(P_i^*(k)) | i \in Z_{[1,L]}, k \in R\}$$

$$\psi^*_{min} = min\{\psi_{min}(P_i^*(k)) | i \in Z_{[1,L]}, k \in R\}$$
where $\psi_{max}(\cdot)$ and $\psi_{min}(\cdot)$ are, respectively, the maximal and minimal eigenvalues.
in addition, (24) implies that,
$V_k(x_i(k+1)) - \bar{V}(x_i(k))$

$$< -\left(x_i^T(k)Qx_i(k) + u_i^T(k)Ru_i(k) - \tau_i d_i^T(k)d_i(k) + \delta_i^T(k)\left(z_i(k) - \sum_{\substack{j=1\\i\neq j}}^{N} f_{ij}x_j(k)\right)\right.$$

$$\left.+ \bar{p}_i^T(k+1)\left(-x_i(k+1) + g_i(x_i(k), u_i(k), z_i(k))\right)\right) \tag{32}$$

where $V_k(x_i(k+1)) = x_i^T(k+1)P^*_{i\mu}x_i(k+1)$. And if
$$V_k(x_i(k+1)) - V(k, x_i) < -x_i^T(k)Qx_i(k) + \tau_i d_i^T(k)d_i(k) \tag{33}$$
due to the (30) at time $(k+1)$, it will be:
$$V_{k+1}(x_i(k+1)) \leq V_k(x_i(k+1)) \tag{34}$$
finally, it achieves that
$$V_{k+1}(x_i(k+1)) - V(k, x_i) < -x_i^T(k)Qx_i(k) + \tau_i d_i^T(k)d_i(k) \tag{35}$$

corresponding definition (3), *Lemma 1*, (31), and (35), the $V(x_i(k))$ is an ISS Lyapunov function. And the closed-loop system is ISS due to disturbances. So the proof is complete. ∎

IV. NUMERICAL EXAMPLE

In this section, to prove the effectiveness of the proposed algorithm, two examples are illustrated, the mentioned algorithm is applied, and results with explanations are depicted. Consider a large-scale system $S$ composed of three fuzzy subsystems $S_i$, $i = 1,2,3$ ($N = 3$), as follows, in which each state of each subsystem has two dimensions. Let $\varpi_i = 0.5$, $Q = diag\{5,5\}$, $H_i = 5$. In this example, the delay time is considered as 1. The membership function, $\mu_i^1 = cos^2(x_{i2}(k))$, $\mu_i^2 = 1 - \mu_i^1$. The T-S fuzzy large-scale model with time-varying delay is:

$$\begin{cases} \text{IF } z_{i1} \text{ is } F_{i1}^l \text{ and } \dots \text{ and } z_{ig} \text{ is } F_{ig}^l \\ \text{THEN } x_i(k+1) = A_i^l x_i(k) + B_i^l u_i(k) + A_{id}^l x_{id}(k) + w_i^l d_i(k) + \sum_{\substack{j=1 \\ i \neq j}}^{N} f_{ij} x_j(k) \end{cases}$$

Subsystem $S_1$

$$A_{11} = \begin{bmatrix} 0.55 & 0.05 \\ 0 & 0.42 \end{bmatrix}, A_{11d} = 0.5 A_{11}, B_{11} = \begin{bmatrix} 1 \\ 0 \end{bmatrix}, w_{11} = \begin{bmatrix} 0.1 \\ 0 \end{bmatrix}, f_{12} = \begin{bmatrix} 0.08 & 0.05 \\ 0.05 & 0.05 \end{bmatrix},$$

$$f_{13} = \begin{bmatrix} 0.09 & 0.06 \\ 0.06 & 0.09 \end{bmatrix}, \lambda_1 = 0.5, X_1 = \begin{bmatrix} 0.015 & 0 \\ 0 & 0.015 \end{bmatrix}$$

$$A_{12} = \begin{bmatrix} 0.4 & 0 \\ 0 & 0.08 \end{bmatrix}, A_{12d} = 0.5 A_{12}, B_{12} = \begin{bmatrix} 0 \\ 1 \end{bmatrix}, w_{12} = \begin{bmatrix} 0 \\ 0.1 \end{bmatrix}.$$

Subsystem $S_2$

$$A_{21} = \begin{bmatrix} 0.325 & 0 \\ 0.4 & 0 \end{bmatrix}, A_{21d} = 0.5 A_{21}, B_{21} = \begin{bmatrix} 1 \\ -1 \end{bmatrix}, w_{21} = \begin{bmatrix} -0.1 \\ 0 \end{bmatrix}, f_{21} = \begin{bmatrix} 0.1 & 0.1 \\ 0 & 0 \end{bmatrix}$$

$$f_{23} = \begin{bmatrix} 0 & 0 \\ 0.1 & 0.1 \end{bmatrix}, \lambda_2 = 0.488, X_2 = \begin{bmatrix} 0.018 & 0 \\ 0 & 0.018 \end{bmatrix}$$

$$A_{22} = \begin{bmatrix} 0.6 & 0.2 \\ 0.1 & 0 \end{bmatrix}, A_{22d} = 0.5 A_{22}, B_{22} = \begin{bmatrix} -1 \\ 1 \end{bmatrix}, w_{22} = \begin{bmatrix} 0 \\ -0.2 \end{bmatrix}.$$

Subsystem $S_3$

$$A_{31} = \begin{bmatrix} 0.2 & 0.4 \\ 0.2 & 0 \end{bmatrix}, A_{31d} = 0.5 A_{31}, B_{31} = \begin{bmatrix} 1 \\ 1 \end{bmatrix}, w_{31} = \begin{bmatrix} -0.3 \\ 0 \end{bmatrix}, f_{31} = \begin{bmatrix} 0.03 & 0 \\ 0 & 0.02 \end{bmatrix},$$

$$f_{32} = \begin{bmatrix} 0.1 & 0 \\ 0.1 & 0 \end{bmatrix}, \lambda_3 = 0.487, X_3 = \begin{bmatrix} 0.027 & 0 \\ 0 & 0.027 \end{bmatrix}$$

$$A_{32} = \begin{bmatrix} 0.3 & 0 \\ 0 & 0.4 \end{bmatrix}, A_{32d} = 0.5 A_{32}, B_{32} = \begin{bmatrix} -2 \\ 1 \end{bmatrix}, w_{32} = \begin{bmatrix} 0 \\ -0.4 \end{bmatrix},$$

where $x_1(k) = [x_{11}(k) \quad x_{12}(k)]^T$, $x_2(k) = [x_{21}(k) \quad x_{22}(k)]^T$, $x_1(k) = [x_{31}(k) \quad x_{32}(k)]^T$, and we have:

$$S_i = x_i(k+1) = \sum_{l=1}^{r_i} \sum_{m=1}^{r_i} \mu_i^l(z_{iq}) \mu_i^m(z_{iq}) [A_{i\mu}^l + B_{i\mu}^l k_i^m] x_i(k) + \mu_i^l(z_{iq}) A_{id\mu}^l x_{id}(k) + w_{i\mu}^l d_i(k)$$

$$+ \sum_{l=1}^{r_i} \sum_{\substack{j=1 \\ i \neq j}}^{N} \mu_i^l(z_{iq}) f_{ij} x_j(k).$$

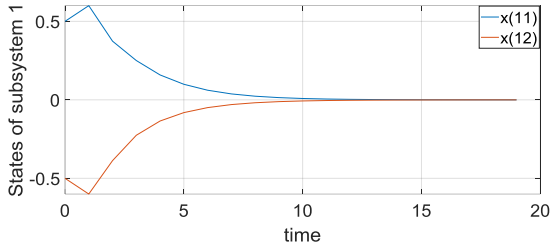

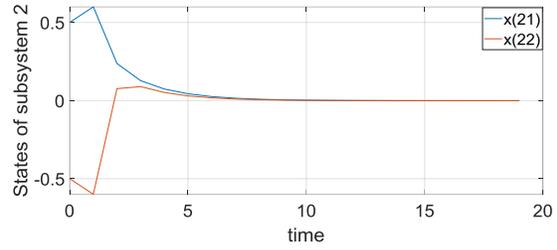

Fig. 2 Trajectories of subsystem 1

Fig. 3 Trajectories of subsystem 2

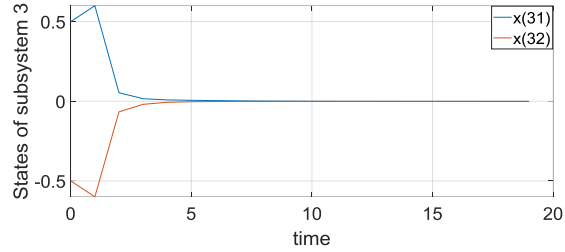

Fig. 4 Trajectories of subsystem 3

*Remark 4:* As it is evident in simulation results, Fig. 2, the trajectories of the subsystem 1, at first, the states of the two rules went to instability due to the existence of the time-varying delay, which is 1 second. But as they go on, the proposed controller has learned to compute efficient gains, which converged the states to 0. Same as Fig. 2, in Fig. 3, and Fig. 4, it is clear that the states of the subsystems were attempting to have overshoot and undershoot. But controllers have successfully stabilized states as time keeps going. The settling time in the first subsystem is before 10, but in subsystems 2 and 3 is before 5, which means the subsystem 1 is more complex than two others.

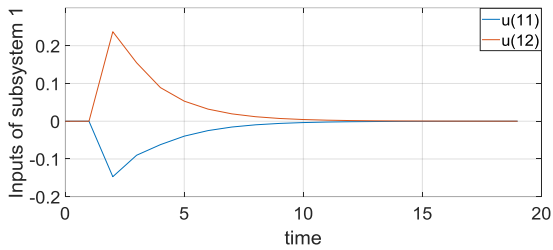

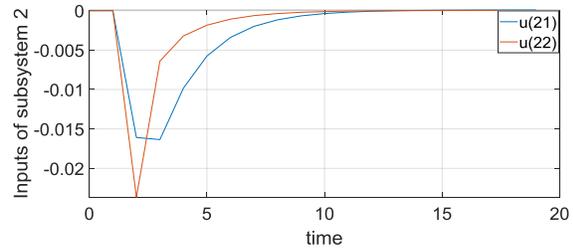

Fig. 5 Trajectories of controller 1

Fig. 6 Trajectories of controller 2

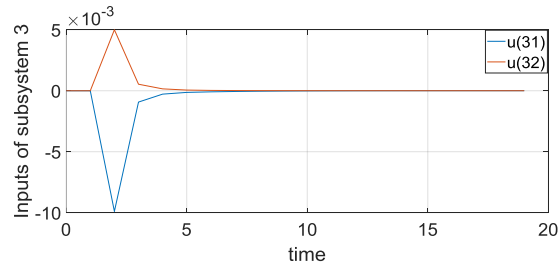

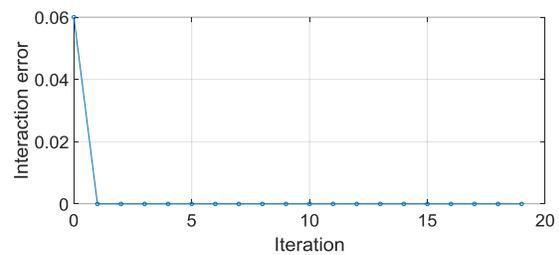

Fig. 7 Trajectories of controller 3

Fig. 8 Interaction error

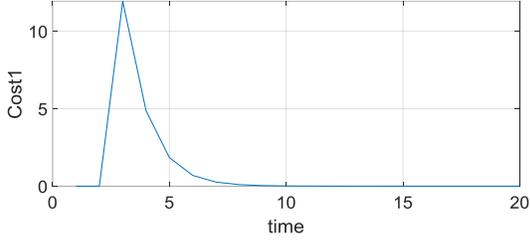
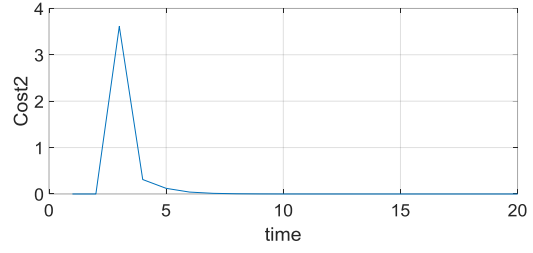

Fig. 9 Trajectory of cost function 1

Fig. 10 Trajectory of cost function 2

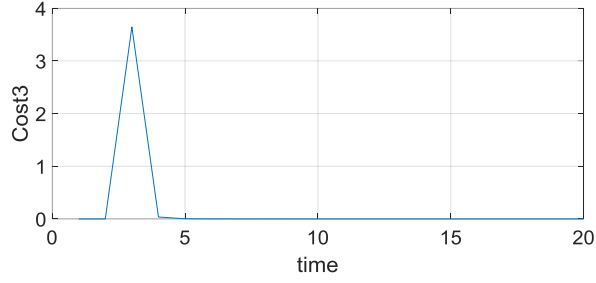

Fig. 11 Trajectory of cost function 3

*Remark 5:* According to Fig. 5, Fig. 6, and Fig. 7, it will be noticeable that the trajectories of the input vector $u_i \in R^n$ are converged to 0. Therefore, there is no need to input vector and costs will be reduced. And also by referring to Fig. 8, after 2 or 3 iterations, the interaction error leads to 0, means that, local subsystems are fully optimized. Finally, Fig. 9, Fig. 10, and Fig. 11, show 3 cost functions corresponding to 3 subsystems and it is obvious that despite of the fact that cost functions involve negative terms, they are definitely positive during the time. It would be necessary to say that the mentioned numerical example is illustrated to show the effectiveness and usefulness of the proposed method, and it will be definitely a practical method in real conditions.

*Remark 6:* In comparison, in [23] and [28], the model predictive control is applied to a usual system with persistent disturbances and uncertainties. But a noticeable issue is that the considered system is a usual one and the proposed algorithm is not efficient for large-scale systems. On the other hand, the mentioned paper has not considered the time-varying delay that is an inevitable part in real systems. In [21], an algorithm is proposed for typical systems with time-varying delay. The model predictive control is assumed for the mentioned research but as mentioned above, it is not considered for large-scale systems. In this research, model predictive control is applied to a fuzzy large-scale system with time-varying delay and persistence disturbances, which the proposed algorithm covers the weaknesses of the mentioned researches.

*Example 2:* A double inverted pendulum is considered due to the [38] to show the effectiveness of the method. All configurations and parameters are chosen same as [38], but a 0.5 sec time-varying delay in considered for the system. All other configurations and considered assumptions are same as the previous example.

Subsystem $S_1$

$A_{11} = A_{13} = \begin{bmatrix} 1 & 0.005 \\ 0.0262 & 1 \end{bmatrix}, A_{13} = \begin{bmatrix} 1 & 0.005 \\ 0.0441 & 1 \end{bmatrix}, A_{11d} = 0.5A_{11}, A_{12d} = 0.5A_{12}, A_{13d} = 0.5A_{13}, B_{11} = B_{12} = B_{13} = \begin{bmatrix} 1 \\ 0 \end{bmatrix}, w_{11} = w_{12} = w_{13} = \begin{bmatrix} 0.1 \\ 0 \end{bmatrix}, g_{12} = \begin{bmatrix} 0.08 & 0.05 \\ 0.05 & 0.05 \end{bmatrix}, \lambda_1 = 0.5, X_1 = \begin{bmatrix} 0.015 & 0 \\ 0 & 0.015 \end{bmatrix}, H_1 = [1\ 0].$

Subsystem $S_2$

$A_{21} = A_{23} = \begin{bmatrix} 1 & 0.005 \\ 0.0272 & 1 \end{bmatrix}, A_{23} = \begin{bmatrix} 1 & 0.005 \\ 0.0451 & 1 \end{bmatrix}, A_{21d} = 0.5A_{21}, A_{22d} = 0.5A_{22}, A_{23d} = 0.5A_{23}, B_{21} = B_{22} = B_{23} = \begin{bmatrix} 1 \\ 1 \end{bmatrix}, w_{11} = w_{22} = w_{23} = \begin{bmatrix} 0.1 \\ 0 \end{bmatrix}, g_{21} = \begin{bmatrix} 0.08 & 0.05 \\ 0.05 & 0.05 \end{bmatrix}, \lambda_2 = 0.448, X_2 = \begin{bmatrix} 0.018 & 0 \\ 0 & 0.018 \end{bmatrix}, H_2 = [1\ 0].$

$$x_i(k+1) = \sum_{l=1}^{r_i}\sum_{m=1}^{r_i}\mu_i^l(z_{iq})\mu_i^m(z_{iq})[A_{i\mu}^l + B_{i\mu}^l k_i^m]x_i(k) + \mu_i^l(z_{iq})A_{id\mu}^l x_{id}(k) + w_{i\mu}^l d_i(k)$$
$$+ \sum_{l=1}^{r_i}\sum_{\substack{j=1\\i\neq j}}^{N}\mu_i^l(z_{iq})f_{ij}x_j(k).$$

$$y_i(k) = H_i x_i(k)$$

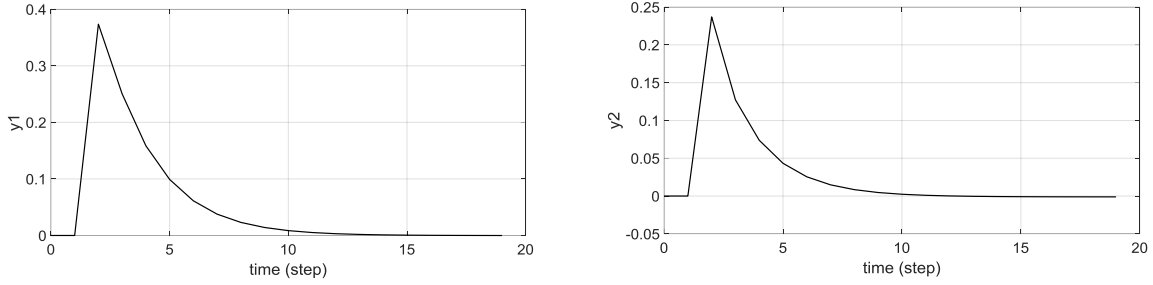

Fig. 12 Output responses of the closed-loop system

Subsequently, we can obtain the feedback gains:

k11= [-4.54  -6.06]   k12= [-6.009  -8.79]   k13= [-15.15 -19.585]

k21= [-5.14  -3.01]   k22= [-1.049  -4.14]   k23= [-28.255 -12.252]

*Remark 7:* Fig. 12 shows output responses of the closed-loop discrete-time nonlinear large-scale system with time-varying delay. It is evident that the proposed controller in this paper based on the fuzzy dynamic model not only stabilizes the original nonlinear large-scale system but also effectively attenuates the disturbances as expected. And as it is clear, for the first few seconds there is no response due to the delay that is in the nature of the system.

*Remark 8:* To prove the effectiveness of the proposed method, a comparison is made due to [39]. In [39], the PI controller is considered for a large-scale system. At first, a control scheme using PI controllers, which are currently used in industry, is used. although this scheme is very simple to implement, its performance has often limitation and tuning the P and I gains are a tedious process. Besides, the mentioned controlled is applied to a system without time-varying delay. Todays, many systems in industry and academic face many delays in their natures and the proposed method in [39] is not able to stabilize the system.

*Remark 9:* By referring to gains, a notable point in this example is results of gains. To stabilize the system with time-varying delay, the computed gains have little values and this means the double inverted pendulum can be stabilized with lower cost and this is the efficient of the proposed approach.

## V. CONCLUSION

Here, fuzzy model predictive control and hierarchical optimization for a class of discrete large-scale systems with time-varying delay and disturbance is investigated. The considered method in this paper is Razumikhin for time-varying delay systems, in which it includes a Lyapunov function associated with the original non-augmented state space of system dynamics. The Razumikhin method has the perfect potential to avoid the inherent complexity of the Krasovskii method especially in the presence of large delays and disturbances. Model predictive control is applied to the fuzzy Takagi-Sugeno large-scale system, and the system is composed of several subsystems to consider the decentralized scheme for controller. Many similar methods have been done before in this area and Razumikhin approach was applied to. But the novelty of this paper is that this method is applied to a large-scale system and simultaneously the optimization problem is considered. On the other hand, by robust performance and LMI algorithm, the stability of the results is guaranteed. In this paper, the uncertainties and disturbances that are always seen in systems are considered. As the assumed system in this paper is large-scale and there are some subsystems, the hierarchical approach is applied to the optimization problem which divides the system into two upper and lower orders. At the upper order, coordination, and at the lower order, some units are established. The exchange information between two orders, finally, leads to the most efficient in the optimization problem. This algorithm is assumed for systems with known time-varying delay systems and in future works it can be applied to nonlinear dynamics with unknown time-varying delay. At least two examples are submitted to investigate the effectiveness of the proposed method.

APPENDIX **A** (PROOF OF THEOREM 1)

***Proof***: Applying Schur complement, the inequality (16) can be written as:

$$\begin{bmatrix} w_{i\mu}^T X_i w_{i\mu} - \varsigma_i \lambda_i \varpi_i & \star & \star & \star & \cdots & \star \\ \Theta_i^T X_i w_{i\mu} & \varphi_i & \star & \star & \cdots & \star \\ A_{id\mu}^T X_i w_{i\mu} & A_{id\mu}^T X_i \Theta_i & A_{id\mu}^T X_i A_{id\mu} - (-\lambda_i + 1)\varrho_{id} X_{i\mu} & \star & \cdots & \star \\ f_{ij}^T X_i w_{i\mu} & (1 - \sqrt{\alpha}) f_{ij}^T X_i \Theta_i & f_{ij}^T X_i A_{id\mu} & -(\alpha - 1) f_{ij}^T X_i f_{ij} & \cdots & \star \\ \vdots & \vdots & \vdots & \vdots & \ddots & \vdots \\ f_{iN}^T X_i w_{i\mu} & (1 - \sqrt{\alpha}) f_{iN}^T X_i \Theta_i & f_{iN}^T X_i A_{id\mu} & -(\alpha - 1) f_{iN}^T X_i f_{ij} & \cdots & -(\alpha - 1) f_{iN}^T X_i f_{iN} \end{bmatrix}$$
$$\leq 0 \qquad (A1)$$

where $\varphi_i = N\Theta_i^T X_i \Theta_i + N\sqrt{\alpha} \sum_{\substack{j=1 \\ i \neq j}}^{N} f_{ij}^T X_j f_{ij} - \varrho_i(-\lambda_i + 1)X_{i\mu}$. By considering $X_i = \varsigma_i P_i, X_{i\mu} = \varsigma_i P_{i\mu}, X_j = \varsigma_i P_j$, $\varpi_i = \frac{\varsigma_i}{\gamma_i^2}$, now, the inequality (A1) is:

$$\begin{bmatrix} \frac{1}{\varsigma_i} w_{i\mu}^T P_i w_{i\mu} - \frac{\lambda_i}{\gamma_i^2} & \star & \star & \star & \cdots & \star \\ \frac{1}{\varsigma_i} \Theta_i^T P_i w_{i\mu} & \varphi_i & \star & \star & \cdots & \star \\ \frac{1}{\varsigma_i} A_{id\mu}^T P_i w_{i\mu} & \frac{1}{\varsigma_i} A_{id\mu}^T P_i \Theta_i & \frac{1}{\varsigma_i} A_{id\mu}^T P_i A_{id\mu} - \left(\frac{-\lambda_i + 1}{\varsigma_i}\right) \varrho_{id} P_{i\mu} & \star & \cdots & \star \\ \frac{1}{\varsigma_i} f_{ij}^T P_i w_{i\mu} & (1 - \sqrt{\alpha}) \frac{1}{\varsigma_i} f_{ij}^T P_i \Theta_i & \frac{1}{\varsigma_i} f_{ij}^T P_i A_{id\mu} & -\frac{1}{\varsigma_i} (\alpha - 1) f_{ij}^T P_i f_{ij} & \cdots & \star \\ \vdots & \vdots & \vdots & \vdots & \ddots & \vdots \\ \frac{1}{\varsigma_i} f_{iN}^T P_i w_{i\mu} & (1 - \sqrt{\alpha}) \frac{1}{\varsigma_i} f_{iN}^T P_i \Theta_i & \frac{1}{\varsigma_i} f_{iN}^T P_i A_{id\mu} & -\frac{1}{\varsigma_i} (\alpha - 1) f_{iN}^T P_i f_{ij} & \cdots & -\frac{1}{\varsigma_i} (\alpha - 1) f_{iN}^T P_i f_{iN} \end{bmatrix}$$
$$\leq 0 \qquad (A2)$$

$\varphi_i = \frac{1}{\varsigma_i} N\Theta_i^T P_i \Theta_i + \frac{1}{\varsigma_i} N\sqrt{\alpha} \sum_{\substack{j=1 \\ i \neq j}}^{N} f_{ij}^T P_j f_{ij} - \frac{1}{\varsigma_i} \varrho_i(-\lambda_i + 1)P_{i\mu}$, now, if the inequality (A2) multiplied from both sides by $\begin{bmatrix} d_i^T & x_i^T & x_i^T & x_j^T & \cdots & x_N^T \end{bmatrix}$ and its transpose, respectively,

$$x_i^T \frac{1}{\varsigma_i} \left\{ N(A_{i\mu} + B_{i\mu}k_{i\mu})^T P_i(A_{i\mu} + B_{i\mu}k_{i\mu}) + N\sqrt{a} \sum_{\substack{j=1 \\ i \neq j}}^N f_{ij}^T P_j f_{ij} - \varrho_i P_{i\mu} \right\} x_i + x_{id}^T \frac{1}{\varsigma_i} A_{id\mu}^T P_i(A_{i\mu} + B_{i\mu}k_{i\mu}) x_i$$

$$+ d_i^T \frac{1}{\varsigma_i} w_{i\mu}^T P_i(A_{i\mu} + B_{i\mu}k_{i\mu}) x_i + x_i^T \frac{1}{\varsigma_i} (A_{i\mu} + B_{i\mu}k_{i\mu})^T P_i w_{i\mu} d_i + x_{id}^T \frac{1}{\varsigma_i} A_{id\mu}^T P_i w_{i\mu} d_i$$

$$+ d_i^T \frac{1}{\varsigma_i} w_{i\mu}^T P_i w_{i\mu} d_i + \frac{1}{\varsigma_i} \left( \sum_{\substack{j=1 \\ i \neq j}}^N x_j^T f_{ij}^T \right) P_i w_{i\mu} d_i + x_i^T \frac{1}{\varsigma_i} (A_{i\mu} + B_{i\mu}k_{i\mu})^T P_i A_{id\mu} x_{id}$$

$$+ x_{id}^T \frac{1}{\varsigma_i} A_{id\mu}^T P_i A_{id\mu} x_{id} + d_i^T \frac{1}{\varsigma_i} w_{i\mu}^T P_i A_{id\mu} x_{id} + \frac{1}{\varsigma_i} \left( \sum_{\substack{j=1 \\ i \neq j}}^N x_j^T f_{ij}^T \right) P_i A_{id\mu} x_{id}$$

$$+ x_{id}^T \frac{1}{\varsigma_i} A_{id\mu}^T P_i \left( \sum_{\substack{j=1 \\ i \neq j}}^N f_{ij} x_j \right) + d_i^T \frac{1}{\varsigma_i} w_{i\mu}^T P_i \left( \sum_{\substack{j=1 \\ i \neq j}}^N f_{ij} x_j \right) - \frac{1}{\varsigma_i}(\alpha - 1) \left( \sum_{\substack{j=1 \\ i \neq j}}^N x_j^T f_{ij}^T \right) P_i \left( \sum_{\substack{j=1 \\ i \neq j}}^N f_{ij} x_j \right)$$

$$+ x_i^T (1 - \sqrt{\alpha}) \frac{1}{\varsigma_i} (A_{i\mu} + B_{i\mu}k_{i\mu})^T P_i \left( \sum_{\substack{j=1 \\ i \neq j}}^N f_{ij} x_j \right) + (1 - \sqrt{\alpha}) \frac{1}{\varsigma_i} \left( \sum_{\substack{j=1 \\ i \neq j}}^N x_j^T f_{ij}^T \right) P_i (A_{i\mu} + B_{i\mu}k_{i\mu}) x_i$$

$$- \frac{\lambda_i}{\gamma_i^2} d_i^T d_i + \left( \frac{\lambda_i - 1}{\varsigma_i} \right) (\varrho_{id} x_{id}^T P_{i\mu} x_{id}) + \left( \frac{\lambda_i}{\varsigma_i} \right) (\varrho_i x_i^T P_{i\mu} x_i) \leq 0 \tag{A3}$$

according to [40], the inequality (A3) is equivalent to:

$$\frac{1}{\varsigma_i}\left(\left[(A_{i\mu}+B_{i\mu}k_{i\mu})x_i+\sqrt{\alpha}\sum_{\substack{j=1\\i\neq j}}^{N}f_{ij}x_j\right]^T P_i\left[(A_{i\mu}+B_{i\mu}k_{i\mu})x_i+\sqrt{\alpha}\sum_{\substack{j=1\\i\neq j}}^{N}f_{ij}x_j\right]-\varrho_i x_i^T P_{i\mu}x_i\right)$$

$$+x_{id}^T\frac{1}{\varsigma_i}A_{id\mu}^T P_i(A_{i\mu}+B_{i\mu}k_{i\mu})x_i+d_i^T\frac{1}{\varsigma_i}w_{i\mu}^T P_i(A_{i\mu}+B_{i\mu}k_{i\mu})x_i$$

$$+x_i^T\frac{1}{\varsigma_i}(A_{i\mu}+B_{i\mu}k_{i\mu})^T P_i w_{i\mu}d_i+x_{id}^T\frac{1}{\varsigma_i}A_{id\mu}^T P_i w_{i\mu}d_i+d_i^T\frac{1}{\varsigma_i}w_{i\mu}^T P_i w_{i\mu}d_i$$

$$+\frac{1}{\varsigma_i}\left(\sum_{\substack{j=1\\i\neq j}}^{N}x_j^T f_{ij}^T\right)P_i w_{i\mu}d_i+x_i^T\frac{1}{\varsigma_i}(A_{i\mu}+B_{i\mu}k_{i\mu})^T P_i A_{id\mu}x_{id}+x_{id}^T\frac{1}{\varsigma_i}A_{id\mu}^T P_i A_{id\mu}x_{id}$$

$$+d_i^T\frac{1}{\varsigma_i}w_{i\mu}^T P_i A_{id\mu}x_{id}+\frac{1}{\varsigma_i}\left(\sum_{\substack{j=1\\i\neq j}}^{N}x_j^T f_{ij}^T\right)P_i A_{id\mu}x_{id}+x_{id}^T\frac{1}{\varsigma_i}A_{id\mu}^T P_i\left(\sum_{\substack{j=1\\i\neq j}}^{N}f_{ij}x_j\right)$$

$$+d_i^T\frac{1}{\varsigma_i}w_{i\mu}^T P_i\left(\sum_{\substack{j=1\\i\neq j}}^{N}f_{ij}x_j\right)-\frac{1}{\varsigma_i}(\alpha-1)\left(\sum_{\substack{j=1\\i\neq j}}^{N}x_j^T f_{ij}^T\right)P_i\left(\sum_{\substack{j=1\\i\neq j}}^{N}f_{ij}x_j\right)$$

$$+x_i^T(1-\sqrt{\alpha})\frac{1}{\varsigma_i}(A_{i\mu}+B_{i\mu}k_{i\mu})^T P_i\left(\sum_{\substack{j=1\\i\neq j}}^{N}f_{ij}x_j\right)$$

$$+(1-\sqrt{\alpha})\frac{1}{\varsigma_i}\left(\sum_{\substack{j=1\\i\neq j}}^{N}x_j^T f_{ij}^T\right)P_i(A_{i\mu}+B_{i\mu}k_{i\mu})x_i-\frac{\lambda_i}{\gamma_i^2}d_i^T d_i+\left(\frac{\lambda_i-1}{\varsigma_i}\right)(\varrho_{id}x_{id}^T P_{i\mu}x_{id})$$

$$+\left(\frac{\lambda_i}{\varsigma_i}\right)(\varrho_i x_i^T P_{i\mu}x_i)\leq 0 \qquad (A4)$$

The inequality (A4) can be written:

$$\frac{1}{\varsigma_i}\left((A_{i\mu}+B_{i\mu}k_{i\mu})x_i+\sum_{\substack{j=1\\i\neq j}}^{N}f_{ij}x_j\right)^T P_i\left((A_{i\mu}+B_{i\mu}k_{i\mu})x_i+\sum_{\substack{j=1\\i\neq j}}^{N}f_{ij}x_j\right)+x_{id}{}^T\frac{1}{\varsigma_i}A_{id\mu}^T P_i(A_{i\mu}+B_{i\mu}k_{i\mu})x_i$$

$$+d_i^T\frac{1}{\varsigma_i}w_{i\mu}^T P_i(A_{i\mu}+B_{i\mu}k_{i\mu})x_i+x_i^T\frac{1}{\varsigma_i}(A_{i\mu}+B_{i\mu}k_{i\mu})^T P_i w_{i\mu}d_i+x_{id}{}^T\frac{1}{\varsigma_i}A_{id\mu}^T P_i w_{i\mu}d_i$$

$$+d_i^T\frac{1}{\varsigma_i}w_{i\mu}^T P_i w_{i\mu}d_i+\frac{1}{\varsigma_i}\left(\sum_{\substack{j=1\\i\neq j}}^{N}x_j^T f_{ij}^T\right)P_i w_{i\mu}d_i+x_i^T\frac{1}{\varsigma_i}(A_{i\mu}+B_{i\mu}k_{i\mu})^T P_i A_{id\mu}x_{id}$$

$$+x_{id}{}^T\frac{1}{\varsigma_i}A_{id\mu}^T P_i A_{id\mu}x_{id}+d_i^T\frac{1}{\varsigma_i}w_{i\mu}^T P_i A_{id\mu}x_{id}+\frac{1}{\varsigma_i}\left(\sum_{\substack{j=1\\i\neq j}}^{N}x_j^T f_{ij}^T\right)P_i A_{id\mu}x_{id}$$

$$+x_{id}{}^T\frac{1}{\varsigma_i}A_{id\mu}^T P_i\left(\sum_{\substack{j=1\\i\neq j}}^{N}f_{ij}x_j\right)+d_i^T\frac{1}{\varsigma_i}w_{i\mu}^T P_i\left(\sum_{\substack{j=1\\i\neq j}}^{N}f_{ij}x_j\right)$$

$$+\frac{1}{\varsigma_i}(\alpha-1)\left(\sum_{\substack{j=1\\i\neq j}}^{N}x_j^T f_{ij}^T\right)P_i\left(\sum_{\substack{j=1\\i\neq j}}^{N}f_{ij}x_j\right)-\frac{1}{\varsigma_i}(\alpha-1)\left(\sum_{\substack{j=1\\i\neq j}}^{N}x_j^T f_{ij}^T\right)P_i\left(\sum_{\substack{j=1\\i\neq j}}^{N}f_{ij}x_j\right)$$

$$+x_i^T\sqrt{\alpha}\frac{1}{\varsigma_i}(A_{i\mu}+B_{i\mu}k_{i\mu})^T P_i\left(\sum_{\substack{j=1\\i\neq j}}^{N}f_{ij}x_j\right)-x_i^T\sqrt{\alpha}\frac{1}{\varsigma_i}(A_{i\mu}+B_{i\mu}k_{i\mu})^T P_i\left(\sum_{\substack{j=1\\i\neq j}}^{N}f_{ij}x_j\right)$$

$$+\sqrt{\alpha}\frac{1}{\varsigma_i}\left(\sum_{\substack{j=1\\i\neq j}}^{N}x_j^T f_{ij}^T\right)P_i(A_{i\mu}+B_{i\mu}k_{i\mu})x_i-\sqrt{\alpha}\frac{1}{\varsigma_i}\left(\sum_{\substack{j=1\\i\neq j}}^{N}x_j^T f_{ij}^T\right)P_i(A_{i\mu}+B_{i\mu}k_{i\mu})x_i-\frac{\lambda_i}{\gamma_i{}^2}d_i^T d_i$$

$$-\frac{1-\lambda_i}{\varsigma_i}\left(\varrho_i x_i^T P_{i\mu}x_i+\varrho_{id}x_{id}^T P_{i\mu}x_{id}\right)\leq 0 \tag{A5}$$

the inequality (A5) is equivalent to:

$$\frac{1}{\varsigma_i}\left((A_{i\mu}+B_{i\mu}k_{i\mu})x_i+A_{id\mu}x_{id}+w_{i\mu}d_i+\sum_{\substack{j=1\\i\neq j}}^{N}f_{ij}x_j\right)^T P_{i\mu}^+\left((A_{i\mu}+B_{i\mu}k_{i\mu})x_i+A_{id\mu}x_{id}+w_{i\mu}d_i+\sum_{\substack{j=1\\i\neq j}}^{N}f_{ij}x_j\right)$$

$$-\frac{\lambda_i}{\gamma_i{}^2}d_i^T d_i\leq\frac{1-\lambda_i}{\varsigma_i}\left(\varrho_i x_i^T P_{i\mu}x_i+\varrho_{id}x_{id}^T P_{i\mu}x_{id}\right) \tag{A6}$$

the equation $x_i^+=A_{i\mu}x_i+A_{id\mu}x_{id}+B_{i\mu}u_i+w_{i\mu}d_i+\sum_{\substack{j=1\\i\neq j}}^{N}f_{ij}x_j$ is results:

$$\sum_{i=1}^{N}\left\{\frac{1}{\varsigma_i}x_i^{+T}P_{i\mu}^+x_i^+ - \frac{\lambda_i}{\gamma_i^2}d_i^Td_i \le \frac{1-\lambda_i}{\varsigma_i}\left(\varrho_i x_i^T P_{i\mu}x_i + \varrho_{id}x_{id}^T P_{i\mu}x_{id}\right)\right\} \tag{A7}$$

it is assumed that $\varrho_i + \varrho_{id} = 1$, so it will be easily confirmed that $\varrho_i x_i^T P_{i\mu}x_i + \varrho_{id}x_{id}^T P_{i\mu}x_{id} \le max\{x_i^T P_{i\mu}x_i, x_{id}^T P_{i\mu}x_{id}\}$. Substituting into (A7), thus, (15) is obtained. Furthermore, the input constraint can be acknowledged by (17), and the proof is shown here:

By multiplying $diag\{I, x_i\}$ and its transpose from both sides of (17):

$$\begin{bmatrix} x_i^T Z_i x_i & x_i^T k_{i\mu}^T \\ k_{i\mu}x_i & 1 \end{bmatrix} \ge 0 \tag{A8}$$

applying Schur complement to (A8), then,

$$x_i^T Z_i x_i - (k_{i\mu}x_i)^T(k_{i\mu}x_i) \ge 0 \tag{A9}$$

as $u_i = k_{i\mu}x_i$, the following is got:

$$(u_i)^T(u_i) \le x_i^T Z_i x_i = H_i = positive\ value \tag{A10}$$

thus $u_i^T u_i \le H_i$. The proof is, thereby, completed. ∎

APPENDIX **B** (PROOF OF THEOREM 2)

*Proof:* Applying schur compliment to the inequality (36):

$$\begin{bmatrix} w_{i\mu}^T X_i w_{i\mu} - \varsigma_i \tau_i & \star & \star \\ \theta_i^T X_i w_{i\mu} & \chi_i & \star \\ A_{id\mu}^T X_i w_{i\mu} & A_{id\mu}^T X_i \theta_i & A_{id\mu}^T X_i A_{id\mu} - \varrho_{id} X_{i\mu} \\ f_{ij}^T X_i w_{i\mu} & (1-\sqrt{\alpha})f_{ij}^T X_i \theta_i & f_{ij}^T X_i A_{id\mu} \\ \vdots & \vdots & \vdots \\ f_{iN}^T X_i w_{i\mu} & (1-\sqrt{\alpha})f_{iN}^T X_i \theta_i & f_{iN}^T X_i A_{id\mu} \\ 0 & 0 & 0 \end{bmatrix}$$

$$\begin{matrix} \star & \cdots & \star & \star \\ \star & \cdots & \star & \star \\ \star & \cdots & \star & \star \\ -(\alpha-1)f_{ij}^T X_i f_{ij} & \cdots & \star & \star \\ \vdots & \ddots & \vdots & \vdots \\ -(\alpha-1)f_{iN}^T X_i f_{ij} & \cdots & -(\alpha-1)f_{iN}^T X_i f_{iN} & \star \\ -\frac{1}{2}\left(\bar{X}_i^T(k+1) + \varsigma_i \delta_j^T(k)\right)f_{ij} & \cdots & -\frac{1}{2}\left(\bar{X}_i^T(k+1) + \varsigma_i \delta_j^T(k)\right)f_{iN} & \left(\varsigma_i \delta_i^T(k)z_i(k) + \bar{X}_i^T(k+1)z_i(k)\right) \end{matrix}$$
$$< 0 \tag{B1}$$

where $\chi_i = N\theta_i^T X_i \theta_i + N\sqrt{\alpha}\sum_{\substack{j=1\\i\ne j}}^{N} f_{ij}^T X_j f_{ij} - \varrho_i X_{i\mu} + \varsigma_i Q + k_{i\mu}^T H_i k_{i\mu}$. The inequality (B1) is equivalent to:

$$\begin{bmatrix} w_{i\mu}^T P_i w_{i\mu} - \tau_i & \star & \star \\ \theta_i^T P_i w_{i\mu} & \Xi_i & \star \\ A_{id\mu}^T P_i w_{i\mu} & A_{id\mu}^T P_i \theta_i & A_{id\mu}^T P_i A_{id\mu} - \varrho_{id} P_{i\mu} \\ f_{ij}^T P_i w_{i\mu} & (1-\sqrt{\alpha})f_{ij}^T P_i \theta_i & f_{ij}^T P_i A_{id\mu} \\ \vdots & \vdots & \vdots \\ f_{iN}^T P_i w_{i\mu} & (1-\sqrt{\alpha})f_{iN}^T P_i \theta_i & f_{iN}^T P_i A_{id\mu} \\ 0 & 0 & 0 \end{bmatrix}$$

$$\left.\begin{matrix} \star & \cdots & \star & & \star \\ \star & \cdots & \star & & \star \\ \star & \cdots & \star & & \star \\ -(\alpha-1)f_{ij}^T P_i f_{ij} & \cdots & \star & & \star \\ \vdots & \ddots & \vdots & & \vdots \\ -(\alpha-1)f_{iN}^T P_i f_{ij} & \cdots & -(\alpha-1)f_{iN}^T P_i f_{iN} & & \star \\ -\frac{1}{2}\left(\bar{p}_i^T(k+1)+\delta_j^T(k)\right)f_{ij} & \cdots & -\frac{1}{2}\left(\bar{p}_i^T(k+1)+\delta_j^T(k)\right)f_{iN} & \left(\delta_i^T(k)z_i(k)+\bar{p}_i^T(k+1)z_i(k)\right) \end{matrix}\right] < 0 \quad (B2)$$

where $\Xi_i = N\theta_i^T P_i \theta_i + N\sqrt{\alpha}\sum_{\substack{j=1 \\ i\neq j}}^{N} f_{ij}^T P_j f_{ij} - \varrho_i P_{i\mu} + Q + k_{i\mu}^T R k_{i\mu}$.

by multiplying $\begin{bmatrix} d_i^T & x_i^T & x_{id}^T & x_j^T & \cdots & x_N^T & I_i^T \end{bmatrix}$ and its transpose from both sides of the matrix in the inequality (B2), respectively, the following inequality is obtained:

$$\sum_{i=1}^{N} \left\{ x_i^T \left\{ N(A_{i\mu}+B_{i\mu}k_{i\mu})^T P_i(A_{i\mu}+B_{i\mu}k_{i\mu}) + N\sqrt{\alpha} \sum_{\substack{j=1 \\ i\neq j}}^{N} f_{ij}^T P_j f_{ij} - \varrho_i P_{i\mu} \right\} x_i - \varrho_{id} x_{id}^T P_{i\mu} x_{id} \right.$$

$$+ x_{id}^T A_{id\mu}^T P_i (A_{i\mu}+B_{i\mu}k_{i\mu}) x_i + d_i^T w_{i\mu}^T P_i (A_{i\mu}+B_{i\mu}k_{i\mu}) x_i + x_i^T (A_{i\mu}+B_{i\mu}k_{i\mu})^T P_i w_{i\mu} d_i$$

$$+ x_{id}^T A_{id\mu}^T P_i w_{i\mu} d_i + d_i^T w_{i\mu}^T P_i w_{i\mu} d_i + \left( \sum_{\substack{j=1 \\ i\neq j}}^{N} x_j^T f_{ij}^T \right) P_i w_{i\mu} d_i + x_i^T (A_{i\mu}+B_{i\mu}k_{i\mu})^T P_i A_{id\mu} x_{id}$$

$$+ x_{id}^T A_{id\mu}^T P_i A_{id\mu} x_{id} + d_i^T w_{i\mu}^T P_i A_{id\mu} x_{id} + \left( \sum_{\substack{j=1 \\ i\neq j}}^{N} x_j^T f_{ij}^T \right) P_i A_{id\mu} x_{id} + x_{id}^T A_{id\mu}^T P_i \left( \sum_{\substack{j=1 \\ i\neq j}}^{N} f_{ij} x_j \right)$$

$$+ d_i^T w_{i\mu}^T P_i \left( \sum_{\substack{j=1 \\ i\neq j}}^{N} f_{ij} x_j \right) - (\alpha-1) \left( \sum_{\substack{j=1 \\ i\neq j}}^{N} x_j^T f_{ij}^T \right) P_i \left( \sum_{\substack{j=1 \\ i\neq j}}^{N} f_{ij} x_j \right)$$

$$+ x_i^T (1-\sqrt{\alpha})(A_{i\mu}+B_{i\mu}k_{i\mu})^T P_i \left( \sum_{\substack{j=1 \\ i\neq j}}^{N} f_{ij} x_j \right) + (1-\sqrt{\alpha}) \left( \sum_{\substack{j=1 \\ i\neq j}}^{N} x_j^T f_{ij}^T \right) P_i (A_{i\mu}+B_{i\mu}k_{i\mu}) x_i$$

$$+ x_i^T Q x_i + u_i^T k_{i\mu}^T R k_{i\mu} u_i - \tau_i d_i^T d_i + I_i^T \left( \delta_i^T(k) z_i(k) + \bar{p}_i^T(k+1) z_i(k) \right) I_i$$

$$\left. - \frac{1}{2} \sum_{\substack{j=1 \\ i\neq j}}^{N} I_i^T \left( \bar{p}_i^T(k+1) + \delta_j^T(k) \right) f_{ij} x_j - \frac{1}{2} \sum_{\substack{j=1 \\ i\neq j}}^{N} x_j^T f_{ij}^T \left( \bar{p}_i^T(k+1) + \delta_j^T(k) \right)^T I_i \right\} < 0 \quad (B3)$$

according to [40], the inequality (B3) is equivalent to:

$$\sum_{i=1}^{N}\left\{\left(\left[(A_{i\mu}+B_{i\mu}k_{i\mu})x_i+\sqrt{\alpha}\sum_{\substack{j=1\\i\neq j}}^{N}f_{ij}x_j\right]^T P_{i\mu}^+\left[(A_{i\mu}+B_{i\mu}k_{i\mu})x_i+\sqrt{\alpha}\sum_{\substack{j=1\\i\neq j}}^{N}f_{ij}x_j\right]-\varrho_i x_i^T P_{i\mu}x_i\right)-\varrho_{id}x_{id}^T P_{i\mu}x_{id}\right.$$

$$+x_{id}{}^T A_{id\mu}^T P_i(A_{i\mu}+B_{i\mu}k_{i\mu})x_i+d_i^T w_{i\mu}^T P_i(A_{i\mu}+B_{i\mu}k_{i\mu})x_i+x_i^T(A_{i\mu}+B_{i\mu}k_{i\mu})^T P_i w_{i\mu}d_i$$

$$+x_{id}{}^T A_{id\mu}^T P_i w_{i\mu}d_i+d_i^T w_{i\mu}^T P_i w_{i\mu}d_i+\left(\sum_{\substack{j=1\\i\neq j}}^{N}x_j^T f_{ij}^T\right)P_i w_{i\mu}d_i+x_i^T(A_{i\mu}+B_{i\mu}k_{i\mu})^T P_i A_{id\mu}x_{id}$$

$$+x_{id}{}^T A_{id\mu}^T P_i A_{id\mu}x_{id}+d_i^T w_{i\mu}^T P_i A_{id\mu}x_{id}+\left(\sum_{\substack{j=1\\i\neq j}}^{N}x_j^T f_{ij}^T\right)P_i A_{id\mu}x_{id}+x_{id}{}^T A_{id\mu}^T P_i\left(\sum_{\substack{j=1\\i\neq j}}^{N}f_{ij}x_j\right)$$

$$+d_i^T\frac{1}{\varsigma_i}w_{i\mu}^T P_i\left(\sum_{\substack{j=1\\i\neq j}}^{N}f_{ij}x_j\right)-(\alpha-1)\left(\sum_{\substack{j=1\\i\neq j}}^{N}x_j^T f_{ij}^T\right)P_i\left(\sum_{\substack{j=1\\i\neq j}}^{N}f_{ij}x_j\right)$$

$$+x_i^T(1-\sqrt{\alpha})(A_{i\mu}+B_{i\mu}k_{i\mu})^T P_i\left(\sum_{\substack{j=1\\i\neq j}}^{N}f_{ij}x_j\right)+(1-\sqrt{\alpha})\left(\sum_{\substack{j=1\\i\neq j}}^{N}x_j^T f_{ij}^T\right)P_i(A_{i\mu}+B_{i\mu}k_{i\mu})x_i$$

$$+x_i^T Q x_i+u_i^T k_{i\mu}^T R k_{i\mu}u_i-\tau_i d_i^T d_i+I_i^T\left(\delta_i^T(k)z_i(k)+\bar{p}_i^T(k+1)z_i(k)\right)I_i$$

$$\left.-\frac{1}{2}\sum_{\substack{j=1\\i\neq j}}^{N}I_i^T\left(\bar{p}_i^T(k+1)+\delta_j^T(k)\right)f_{ij}x_j-\frac{1}{2}\sum_{\substack{j=1\\i\neq j}}^{N}x_j^T f_{ij}^T\left(\bar{p}_i^T(k+1)+\delta_j^T(k)\right)^T I_i\right\}<0 \qquad (B4)$$

the inequality (B4) is equivalent to:

$$\sum_{i=1}^{N}\left\{\left[(A_{i\mu}+B_{i\mu}k_{i\mu})x_i+A_{id\mu}x_{id}+w_{i\mu}d_i+\sum_{\substack{j=1\\i\neq j}}^{N}f_{ij}x_j\right]^T P_{i\mu}^+\left[(A_{i\mu}+B_{i\mu}k_{i\mu})x_i+A_{id\mu}x_{id}+w_{i\mu}d_i+\sum_{\substack{j=1\\i\neq j}}^{N}f_{ij}x_j\right]\right.$$

$$-(\varrho_i x_i^T P_{i\mu}x_i+\varrho_{id}x_{id}^T P_{i\mu}x_{id})+x_i^T Q x_i+u_i^T k_{i\mu}^T R k_{i\mu}u_i-\tau_i d_i^T d_i$$

$$+I_i^T\left(\delta_i^T(k)z_i(k)+\bar{p}_i^T(k+1)z_i(k)\right)I_i-\frac{1}{2}\sum_{\substack{j=1\\i\neq j}}^{N}I_i^T\left(\bar{p}_i^T(k+1)+\delta_j^T(k)\right)f_{ij}x_j$$

$$-\frac{1}{2}\sum_{\substack{j=1\\i\neq j}}^{N}x_j^T f_{ij}^T\left(\bar{p}_i^T(k+1)+\delta_j^T(k)\right)^T I_i-(a-1)\sum_{\substack{j=1\\i\neq j}}^{N}x_j^T f_{ij}^T P_i f_{ij}x_j+(a-1)\sum_{\substack{j=1\\i\neq j}}^{N}x_j^T f_{ij}^T P_i f_{ij}x_j$$

$$+x_i^T\sqrt{\alpha}(A_{i\mu}+B_{i\mu}k_{i\mu})^T P_i\left(\sum_{\substack{j=1\\i\neq j}}^{N}f_{ij}x_j\right)-x_i^T\sqrt{\alpha}(A_{i\mu}+B_{i\mu}k_{i\mu})^T P_i\left(\sum_{\substack{j=1\\i\neq j}}^{N}f_{ij}x_j\right)$$

$$\left.+\sqrt{\alpha}\left(\sum_{\substack{j=1\\i\neq j}}^{N}x_j^T f_{ij}^T\right)P_i(A_{i\mu}+B_{i\mu}k_{i\mu})x_i-\sqrt{\alpha}\left(\sum_{\substack{j=1\\i\neq j}}^{N}x_j^T f_{ij}^T\right)P_i(A_{i\mu}+B_{i\mu}k_{i\mu})x_i\right\}<0 \quad (B5)$$

the inequality (B5) can be written as:

$$\sum_{i=1}^{N}\left\{\left[(A_{i\mu}+B_{i\mu}k_{i\mu})x_i+A_{id\mu}x_{id}+w_{i\mu}d_i+\sum_{\substack{j=1\\i\neq j}}^{N}f_{ij}x_j\right]^T P_{i\mu}^+\left[(A_{i\mu}+B_{i\mu}k_{i\mu})x_i+A_{id\mu}x_{id}+w_{i\mu}d_i+\sum_{\substack{j=1\\i\neq j}}^{N}f_{ij}x_j\right]\right.$$

$$-(\varrho_i x_i^T P_{i\mu}x_i+\varrho_{id}x_{id}^T P_{i\mu}x_{id})+x_i^T Q x_i+u_i^T R u_i-\tau_i d_i^T d_i+\delta_i^T(k)z_i(k)-\sum_{\substack{j=1\\i\neq j}}^{N}\delta_j^T(k)f_{ij}x_j$$

$$\left.+\bar{p}_i^T(k+1)\left(-x_i(k+1)+g_i(x_i(k),u_i(k),z_i(k))\right)\right\}<0 \quad (B6)$$

if it is assumed that $\varrho_i+\varrho_{id}=1$, it will be obvious $\varrho_i x_i^T P_{i\mu}x_i+\varrho_{id}x_{id}^T P_{i\mu}x_{id}\leq V(x_i)$, therefore:

where $\sum_{i=1}^{N}V(x_i^+)-V(x_i)<-\sum_{i=1}^{N}H_i(\cdot)$. Thereby, the Proof is completed. ∎

**R**eferences